\documentclass[aps,prb,twocolumn,amsmath,amssymb,groupedaddress,longbibliography
]{revtex4-2}

\usepackage[dvipdfmx]{graphicx}
\usepackage{dcolumn}
\usepackage{bm}
\usepackage{multirow}
\usepackage{braket}
\usepackage{color}
\usepackage{ulem}
\usepackage{booktabs}

\begin{document}

\title{
Nonlinear nonreciprocal transport in antiferromagnets free from spin-orbit coupling
}

\author{Satoru Hayami$^{1,2}$ and Megumi Yatsushiro$^{1,2}$}
\affiliation{
$^1$Department of Applied Physics, the University of Tokyo, Tokyo 113-8656, Japan \\
$^2$Department of Physics, Hokkaido University, Sapporo 060-0810, Japan 
 }

\begin{abstract}
We theoretically propose a realization of a nonlinear nonreciprocal transport in antiferromagnets without relying on the relativistic spin-orbit coupling. 
Through the symmetry and microscopic model analyses, we show that a local spin scalar chirality to induce an asymmetric band modulation becomes a source of a Drude-type nonlinear transport, while 
an electric polarization induced by a collinear spin configuration in a triangle unit leads to a Berry-curvature-dipole-type nonlinear transport. 
We demonstrate that 120$^\circ$ antiferromagnetic ordering on a triangular lattice and a breathing kagome lattice in an external magnetic field are typical examples.  
Our results open a new direction of designing and engineering functional materials with showing rich parity-violating transport phenomena by spontaneous magnetic phase transitions.
\end{abstract}

\maketitle

\section{Introduction}

The study of magnetism has long been the subject of considerable theoretical and experimental interest in condensed matter physics. 
Once a spontaneous magnetic phase transition occurs, a variety of intriguing physical phenomena appear as a consequence of the interplay between charge and spin degrees of freedom in electrons, such as the magnetoelectric effect irrespective of metals and insulators~\cite{kimura2003magnetic,Fiebig0022-3727-38-8-R01,Katsura_PhysRevLett.95.057205,Mostovoy_PhysRevLett.96.067601,KhomskiiPhysics.2.20,tokura2014multiferroics,Hayami_PhysRevB.90.024432,saito2018evidence,thole2018magnetoelectric,Gao_PhysRevB.97.134423,Shitade_PhysRevB.98.020407}. 
In addition, the topological aspect of magnetic orderings has been extensively studied in recent years since the discovery of magnetic skyrmions~\cite{Muhlbauer_2009skyrmion,yu2010real,nagaosa2013topological,hayami2021topological} and magnetic topological insulators~\cite{tokura2019magnetic,otrokov2019prediction,deng2020quantum,bonbien2021topological}. 
These diverse physical properties in magnetic systems provide us a promising possibility for potential applications to functional spintronics devices~\cite{Zutic_RevModPhys.76.323,jungwirth2016antiferromagnetic,Baltz_RevModPhys.90.015005}. 

Toward applications, it is significant to elucidate the conditions to induce physical phenomena at the fundamental level. 
The symmetry aspect provides their macroscopic conditions. 
Meanwhile, fundamental understanding of the microscopic origins is important to realize an efficient bottom-up design of functional magnetic materials.
One of the successes has been achieved in the anomalous Hall effect in magnetic materials~\cite{Nagaosa_RevModPhys.82.1539,smejkal2021anomalous}. 
Although it was originally discussed in ferromagnets with the relativistic spin-orbit coupling (SOC)~\cite{Karplus_PhysRev.95.1154,smit1958spontaneous,Maranzana_PhysRev.160.421,Berger_PhysRevB.2.4559,nozieres1973simple,Jungwirth_PhysRevLett.88.207208}, unconventional mechanisms in collinear~\cite{Solovyev_PhysRevB.55.8060,vsmejkal2020crystal,Shao_PhysRevApplied.15.024057,samanta2020crystal,Naka_PhysRevB.102.075112,Hayami_PhysRevB.103.L180407}, noncollinear~\cite{Tomizawa_PhysRevB.80.100401,Chen_PhysRevLett.112.017205,nakatsuji2015large,Suzuki_PhysRevB.95.094406,Chen_PhysRevB.101.104418,Lux_PhysRevLett.124.096602}, and noncoplanar~\cite{Ohgushi_PhysRevB.62.R6065,Shindou_PhysRevLett.87.116801,taguchi2001spin,Martin_PhysRevLett.101.156402,Neubauer_PhysRevLett.102.186602} antiferromagnets (AFMs) have been clarified based on the Berry curvature~\cite{Loss_PhysRevB.45.13544,Ye_PhysRevLett.83.3737,Haldane_PhysRevLett.93.206602,Xiao_RevModPhys.82.1959,Zhang_PhysRevB.101.024420} and multipoles~\cite{Suzuki_PhysRevB.95.094406,Hayami_PhysRevB.103.L180407}. 
Among them, noncoplanar spin textures give rise to the anomalous (topological) Hall effect even without either the SOC or uniform magnetic moments, which results in broadening the scope of materials~\cite{tatara2002chirality,Shindou_PhysRevLett.87.116801,Martin_PhysRevLett.101.156402,batista2016frustration}.

The rich physics brought about by the anomalous Hall effect leads to a natural extension to a nonlinear nonreciprocal transport beyond the linear response. 
Although the framework and the symmetry argument for the nonreciprocal transport have been developed~\cite{gao2014, Sodemann_PhysRevLett.115.216806,Parker_PhysRevB.99.045121,Xiao_PhysRevB.100.165422,Nandy_PhysRevB.100.195117,tokura2018nonreciprocal, gao2019semiclassical, Yatsushiro_PhysRevB.104.054412, Oiwa_doi:10.7566/JPSJ.91.014701}, 
few studies have focused on the microscopic mechanisms in magnetic systems~\cite{togawa2016symmetry, ishizuka2020anomalous, Holder_PhysRevResearch.2.033100,Fei_PhysRevB.102.035440,Watanabe_PhysRevResearch.2.043081,Watanabe_PhysRevX.11.011001,yatsushiro2021microscopic, PhysRevLett.127.277201, PhysRevLett.127.277202}. 
Especially, the effect of the SOC can be incorporated in most cases, which narrows down  candidate materials.

In the present study, we explore another mechanism of the nonlinear nonreciprocal transport in AFMs without relying on the SOC. 
Based on the symmetry and model analyses, we derive a microscopic essence for Drude-type and Berry-curvature-dipole(BCD)-type nonlinear nonreciprocal transports, which appear under the different conditions: 
the Drude-type nonlinear transport is induced by a noncoplanar spin configuration with the local spin scalar chirality, while the BCD-type one is induced by a collinear spin configuration in a triangle unit with the electric polarization. 
The results are demonstrated by examining three-sublattice 120$^{\circ}$ AFM orderings on a triangular lattice (TL) and a breathing kagome lattice (BKL) in an external magnetic field. 
We also discuss necessary conditions for model parameters to cause these nonlinear conductivities from the microscopic viewpoint. 
Our results indicate that noncollinear and noncoplanar magnetic orderings in frustrated and itinerant magnets with the negligibly small SOC are also potential candidates to bring about rich parity-violating transport phenomena, which will stimulate further exploration of functional magnetic materials.

The rest of this paper is organized as follows. 
In Sec.~\ref{sec: Model}, we present a tight-binding model without the spin-orbit coupling on the TL and BKL, which consists of the hopping, the site-dependent AFM mean field, and an external magnetic field. 
We show the numerical results of the nonreciprocal transport for both lattice systems in Sec.~\ref{sec: Results}. 
We also discuss the important model parameters to induce the Drude-type and BCD-type nonlinear transports. 
Section~\ref{sec: Summary} is devoted to a summary.

\section{Model}
\label{sec: Model}
\begin{figure}[t!]
\begin{center}
\includegraphics[width=1.0 \hsize]{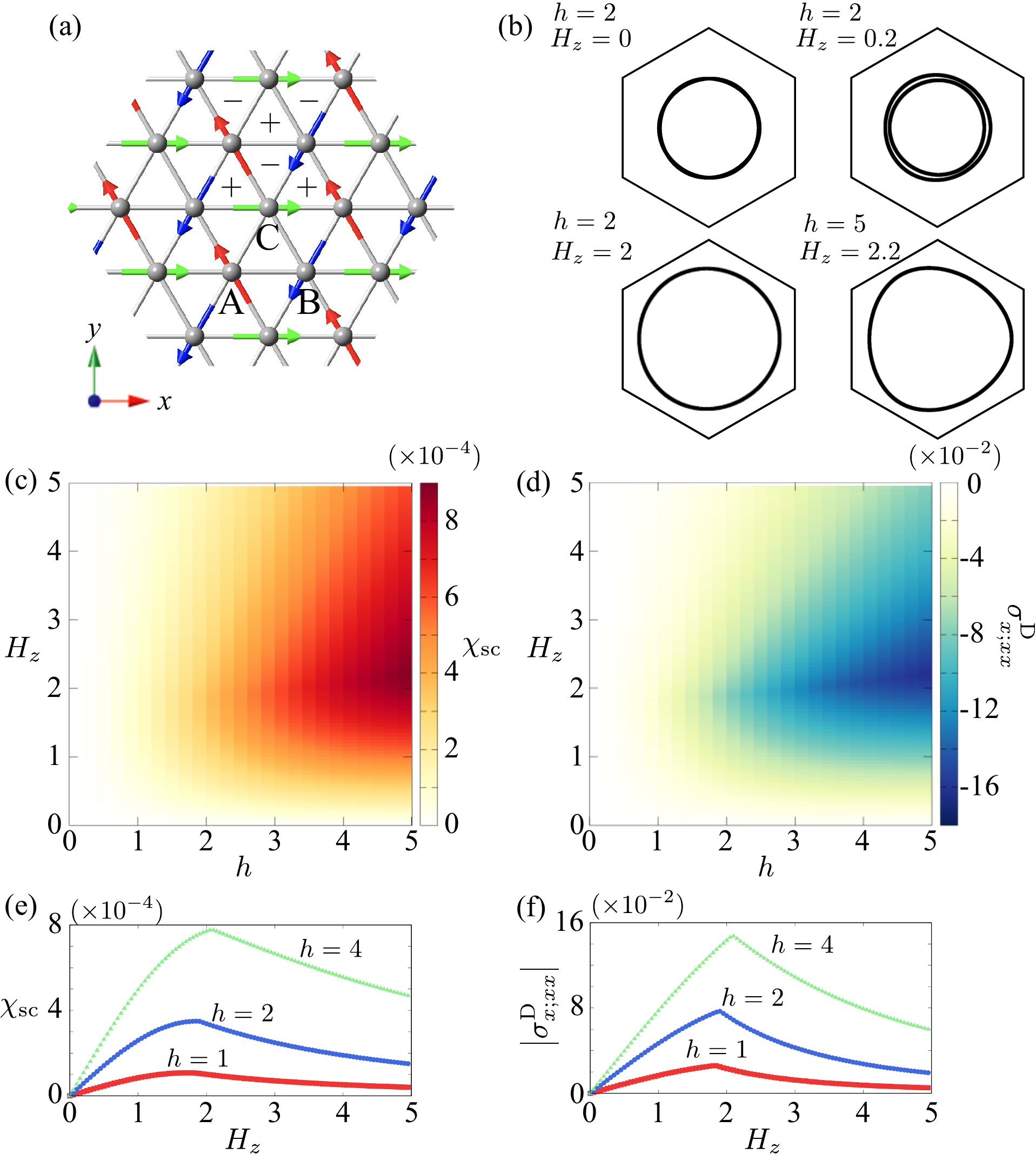} 
\caption{
\label{fig:TL}
(a) 120$^{\circ}$-AFM structure on the TL. 
$+$ or $-$ on the triangle represents the sign of the local spin scalar chirality $\chi_{\rm sc}$ under $H_z$. 
(b) Fermi surfaces at 1/10 filling for several $h$ and $H_z$. 
(c,d) Contour plots of (c) $\chi_{\rm sc}$ and (d) $\sigma^{\rm D}_{x;xx}$ in the plane of $h$ and $H_z$ at 1/10 filling. 
(e,f) $H_z$ dependences of (e) $\chi_{\rm sc}$ and (f) $\sigma^{\rm D}_{x;xx}$ for several $h$. 
}
\end{center}
\end{figure}

Let us start with a tight-binding model on the TL, which is given by 
\begin{align}
\mathcal{H}^{\rm TL}= -t \sum_{\braket{ij}\sigma} c_{i\sigma}^{\dagger}c_{j\sigma}^{} + \sum_{i\sigma \sigma'} (\bm{h}_i+\bm{H}) \cdot 
c_{i\sigma}^{\dagger} \bm{\sigma}_{\sigma \sigma'}c_{i\sigma'}^{},
\label{eq:Ham_TL}
\end{align}
where $c^{\dagger}_{i\sigma}$ ($c_{i\sigma}^{}$) is the creation (annihilation) operator for site $i$ and spin $\sigma=\uparrow, \downarrow$. 
The first term represents the hoppings between the nearest-neighbor sites. 
The second term consists of the site-dependent AFM mean-field term $\bm{h}_i$ that originates from the Coulomb interaction $U$ and an external magnetic field $\bm{H}=(0,H_y,H_z)$ in the $yz$ plane, where $\bm{\sigma}$ is the vector of the Pauli matrices; the magnitude of $\bm{h}_i$ is roughly given by $Um_i$ within the mean-field approximation in the single-band Hubbard model, where $m_i$ represents the expectation value of the spins ($0<m_i\leq0.5$).
For the former, we assume the noncollinear three-sublattice 120$^{\circ}$-AFM structure as $\bm{h}_{\rm A}= h(-1/2, \sqrt{3}/2,0)$, $\bm{h}_{\rm B}= h(-1/2,-\sqrt{3}/2,0)$, and $\bm{h}_{\rm C}= h(1,0,0)$ in Fig.~\ref{fig:TL}(a) with the amplitude $h$ to focus on the emergence of the nonreciprocal transport under noncollinear magnetic orderings.
Such a noncollinear ordering is stabilized in the competing exchange interactions with the triangle unit in the itinerant electron models including not only the Hubbard model but also the periodic Anderson model and the classical Kondo lattice model~\cite{Sahebsara_PhysRevLett.100.136402,Yoshioka_PhysRevLett.103.036401,Yang_PhysRevLett.105.267204,Akagi_JPSJ.79.083711,Hayami2011} or in the magnetic anisotropy.
We set $t=1$ as the energy unit, take the lattice constant as unity, and consider the low 1/10 filling so that the Fermi surfaces consisting of two almost energetically degenerate bands become a simple shape, as shown in Fig.~\ref{fig:TL}(b). 
In addition, to cover the situation from a weak-correlation regime ($U/t \ll 1$) to a strong-correlation regime ($U/t \sim 10$), $h$ is changed from 0 to 5, as discussed in Sec.~\ref{sec: Triangular-lattice system}.

In the BKL system, the Hamiltonian is given by 
\begin{align}
\mathcal{H}^{\rm BKL}=  &-(t_a \sum_{\braket{ij}\sigma}^{\in \triangle} + t_b \sum_{\braket{ij}\sigma}^{\in \bigtriangledown} ) c_{i\sigma}^{\dagger}c_{j\sigma}^{} \nonumber \\
&+ \sum_{i\sigma \sigma'} (\bm{h}_i+\bm{H}) \cdot 
c_{i\sigma}^{\dagger} \bm{\sigma}_{\sigma \sigma'}c_{i\sigma'}^{},
\label{eq:Ham_BKL}
\end{align}
where $t_a$ ($t_b$) represents the hopping within upward (downward) triangles. 
Similar to the TL system, we consider the noncollinear three-sublattice 120$^{\circ}$-AFM structure as $\bm{h}_{\rm A}= h(-1/2, \sqrt{3}/2,0)$, $\bm{h}_{\rm B}= h(-1/2,-\sqrt{3}/2,0)$, and $\bm{h}_{\rm C}= h(1,0,0)$ in Fig.~\ref{fig:BKL_L}(a).
We take the length of both triangles as unity, whose difference is expressed as the different hoppings; $t_a=1$ and $t_b=0.5$. 
The results are shown in Sec.~\ref{sec: Breathing-kagome-lattice system}.

\section{Results}
\label{sec: Results}

We show the results of the TL system in Sec.~\ref{sec: Triangular-lattice system} and those of the BKL system in Sec.~\ref{sec: Breathing-kagome-lattice system}. 
We also discuss the essential model parameters for the Durde-type and BCD-type nonlinear transports in Sec.~\ref{sec: Essential model parameters}. 

\subsection{Triangular-lattice system}
\label{sec: Triangular-lattice system}

The 120$^{\circ}$-AFM structure on the TL breaks the spatial inversion symmetry. 
Then, the antisymmetric spin polarization 
appears in the band structure in the model in Eq.~(\ref{eq:Ham_TL})~\cite{Hayami_PhysRevB.101.220403,Hayami_PhysRevB.102.144441,Yuan_PhysRevMaterials.5.014409}. 
When the magnetic moments are on the $xy$ plane under ${\bm H}={\bm 0}$, the antisymmetric $z$-spin polarization satisfying threefold rotational symmetry occurs in the form of $k_x(k_x^2-3k_y^2)\sigma_z$ ($\bm{k}$ is the wave vector). 
In such a situation, the product symmetry of the sixfold rotation and the space-time-inversion operation remains, which results in the sixfold-symmetric Fermi surface.
When introducing the magnetic field along $\sigma_z$, 
the asymmetry in terms of $+\bm{k}$ and $-\bm{k}$ appears~\cite{Hayami_PhysRevB.101.220403,Hayami_PhysRevB.102.144441}; the sixfold rotational symmetry of the Fermi surface is lost, as shown for several values of $h$ and $H_z$ in Fig.~\ref{fig:TL}(b), where one of the two bands is away from the Fermi level while increasing $H_z$.
Such an asymmetric band structure becomes the microscopic key ingredient to cause the Drude-type nonreciprocal nonlinear transport discussed below.

The Drude-type nonlinear conductivity $\sigma^{\rm D}_{\eta;\mu\nu}$ in $J_{\eta}=\sigma^{\rm D}_{\eta;\mu\nu} E_{\mu}E_{\nu}$ for $\eta,\mu,\nu=x,y,z$ is derived from the second-order Kubo formula as 
\begin{align}
\sigma_{\eta;\mu\nu}^{\rm D}=-\frac{e^3 \tau^2}{2\hbar^3N_{\bm{k}}} \sum_{\bm{k},n}f_{n\bm{k}}\partial_{\eta}\partial_{\mu}\partial_{\nu}\varepsilon_{n\bm{k}}, 
\end{align}
where $e$, $\tau$, $\hbar$, and $N_{\bm k}$ are the electron charge, relaxation time, the reduced Planck constant, and the number of supercells, respectively; we take $e=\tau=\hbar=1$. 
$\varepsilon_{n\bm{k}}$ and $f_{n\bm{k}}$ are the eigenenergy and the Fermi distribution function with the band index $n$, respectively. 
$\sigma_{\eta;\mu\nu}^{\rm D}$ arises for the nonzero $\partial_{\eta}\partial_{\mu}\partial_{\nu}\varepsilon_{n\bm{k}}$, i.e., asymmetrically deformed energy band.
In the present 120$^{\circ}$ AFM in the two-dimentional TL system, $\sigma_{x;xx}^{\rm D}=-\sigma^{\rm D}_{x;yy}=-\sigma^{\rm D}_{y;xy}$ from the symmetry under $H_z$. 
Hereafter, we fix the temperature $T=0.01$ and $N_{\bm{k}}=4800^2$. 

Figures~\ref{fig:TL}(c) and \ref{fig:TL}(d) show the color plots of $\chi_{\rm sc}$ and $\sigma_{x;xx}^{\rm D}$ while changing $h$ and $H_z$ at $H_y=0$, respectively. 
Compared to $\chi_{\rm sc}$ in Fig.~\ref{fig:TL}(c) to $|\sigma_{x;xx}^{\rm D}|$ in Fig.~\ref{fig:TL}(d), one finds that there is a correlation between them in the wide range of $h$ and $H_z$; 
both quantities increase while increasing $h$ at fixed $H_z$, while they become the largest at intermediate $H_z$ at fixed $h$ [Figs.~\ref{fig:TL}(e) and \ref{fig:TL}(f)]. 
This indicates that large $|\sigma_{x;xx}^{\rm D}|$ in AFMs without the SOC can be realized when the noncoplanarity of the spin configuration becomes large.

Let us discuss the relationship among the nonreciprocal transport, local scalar chirality, and the asymmetric band structure for an intuitive understanding. 
The microscopic origin of the nonzero $\sigma_{x;xx}^{\rm D}$ is understood from the distribution of the local spin scalar chirality defined by $\chi_{\rm sc}=\langle \bm{s}_i \rangle \cdot (\langle\bm{s}_j \rangle\times \langle\bm{s}_k\rangle)$ where $\langle \bm{s}_i\rangle =(1/2) \langle\sum_{\sigma\sigma'}c_{i\sigma}^{\dagger} \bm{\sigma}_{\sigma \sigma'}c_{i\sigma'}^{}\rangle$, $\langle \cdots \rangle$ is the expectation value, and $i$, $j$, and $k$ are the sites in the triangle in the counterclockwise order. 
As shown in Fig.~\ref{fig:TL}(a), the 120$^{\circ}$ AFM under $H_z$ leads to the staggered alignment of $\pm \chi_{\rm sc}$ so that the sixfold rotational symmetry of the system is broken in the real-space picture. 
Accordingly, the sixfold symmetry of the Fermi surfaces is also broken, which leads to the asymmetric band deformation. 
The important point is that the uniform component of $\chi_{\rm sc}$ is not necessary to induce nonreciprocal transport; the essence lies in its spatial distribution breaking the sixfold rotational symmetry.  
Thus, the relationship between $\chi_{\rm sc}$ and $\sigma_{x;xx}^{\rm D}$ is qualitatively different from that between $\chi_{\rm sc}$ and linear topological Hall effect, the latter of which requires the uniform alignment of  $\chi_{\rm sc}$.

\begin{figure}[t!]
\begin{center}
\includegraphics[width=1.0 \hsize ]{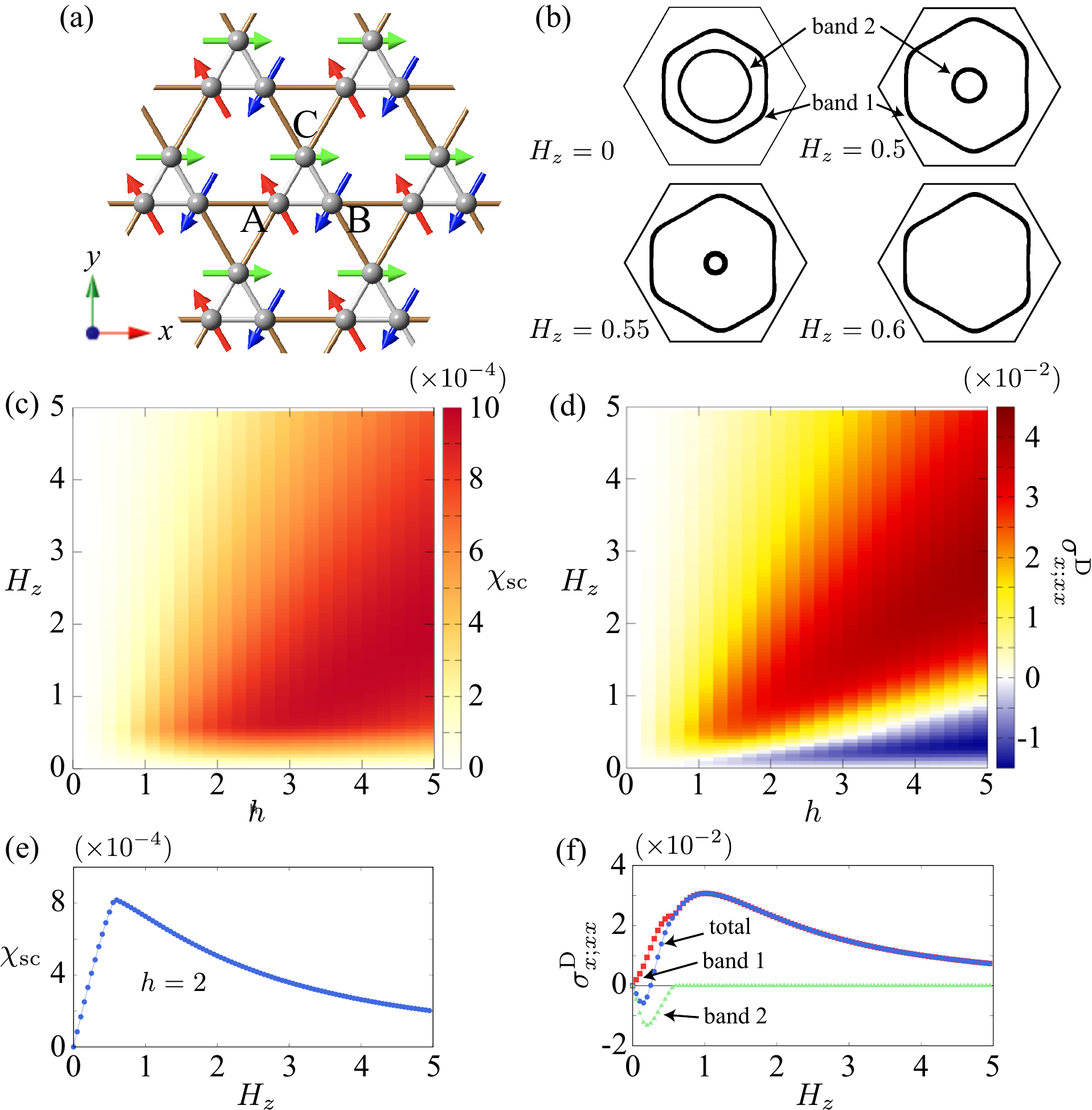} 
\caption{
\label{fig:BKL_L}
(a) 120$^{\circ}$-AFM structure in the BKL system. 
(b) Fermi surfaces at 1/10 filling for several $H_z$ at $h=3$. 
(c,d) Contour plots of (c) $\chi_{\rm sc}$ and (d) $\sigma^{\rm D}_{x;xx}$ in the plane of $h$ and $H_z$ at 1/10 filling. 
(e,f) $H_z$ dependences of (e) $\chi_{\rm sc}$ and (f) $\sigma^{\rm D}_{x;xx}$ for $h=2$. 
In (b,f), the band 1 and band 2 represent the contributions from the lowest and second-lowest bands, respectively.  
}
\end{center}
\end{figure}

\subsection{Breathing-kagome-lattice system}
\label{sec: Breathing-kagome-lattice system}

A similar correlation between $\chi_{\rm sc}$ and $\sigma_{x;xx}^{\rm D}$ also occurs in other noncollinear AFMs. 
To demonstrate that, we investigate the 120$^{\circ}$ AFM ordering in the BKL structure in Fig.~\ref{fig:BKL_L}(a), where the model Hamiltonian is given by Eq.~(\ref{eq:Ham_BKL}). 
The 120$^{\circ}$ AFM ordering in Fig.~\ref{fig:BKL_L}(a) exhibits the asymmetric band deformation in the form of $k_x (k_x^2-3k_y^2)$ when $\chi_{\rm sc}$ is present under $H_z$, as shown by the Fermi surfaces at 1/10 filling and $h=3$ in Fig.~\ref{fig:BKL_L}(b); two bands denoted as the band 1 and band 2 form the Fermi surfaces for small $H_z$ and the band 2 is away from the Fermi level while increasing $H_z$. 

The asymmetric band deformation under nonzero $\chi_{\rm sc}$ leads to nonzero $\sigma_{x;xx}^{\rm D}$. 
Similar to the TL case in Sec.~\ref{sec: Triangular-lattice system}, the behaviors of $\sigma_{x;xx}^{\rm D}$ against $h$ and $H_z$ in Fig.~\ref{fig:BKL_L}(d) have a correspondence to $\chi_{\rm sc}$ in Fig.~\ref{fig:BKL_L}(c) except for the small $H_z$ region [see also the case at $h=2$ in  Figs.~\ref{fig:BKL_L}(e) and \ref{fig:BKL_L}(f)]. 
The deviation for small $H_z$ is owing to the opposite-sign contributions from the two Fermi surfaces, where the negative contribution from the band 2 is larger than the positive one from the band 1, as shown in Fig.~\ref{fig:BKL_L}(f).

\begin{figure}[t!]
\begin{center}
\includegraphics[width=1.0 \hsize ]{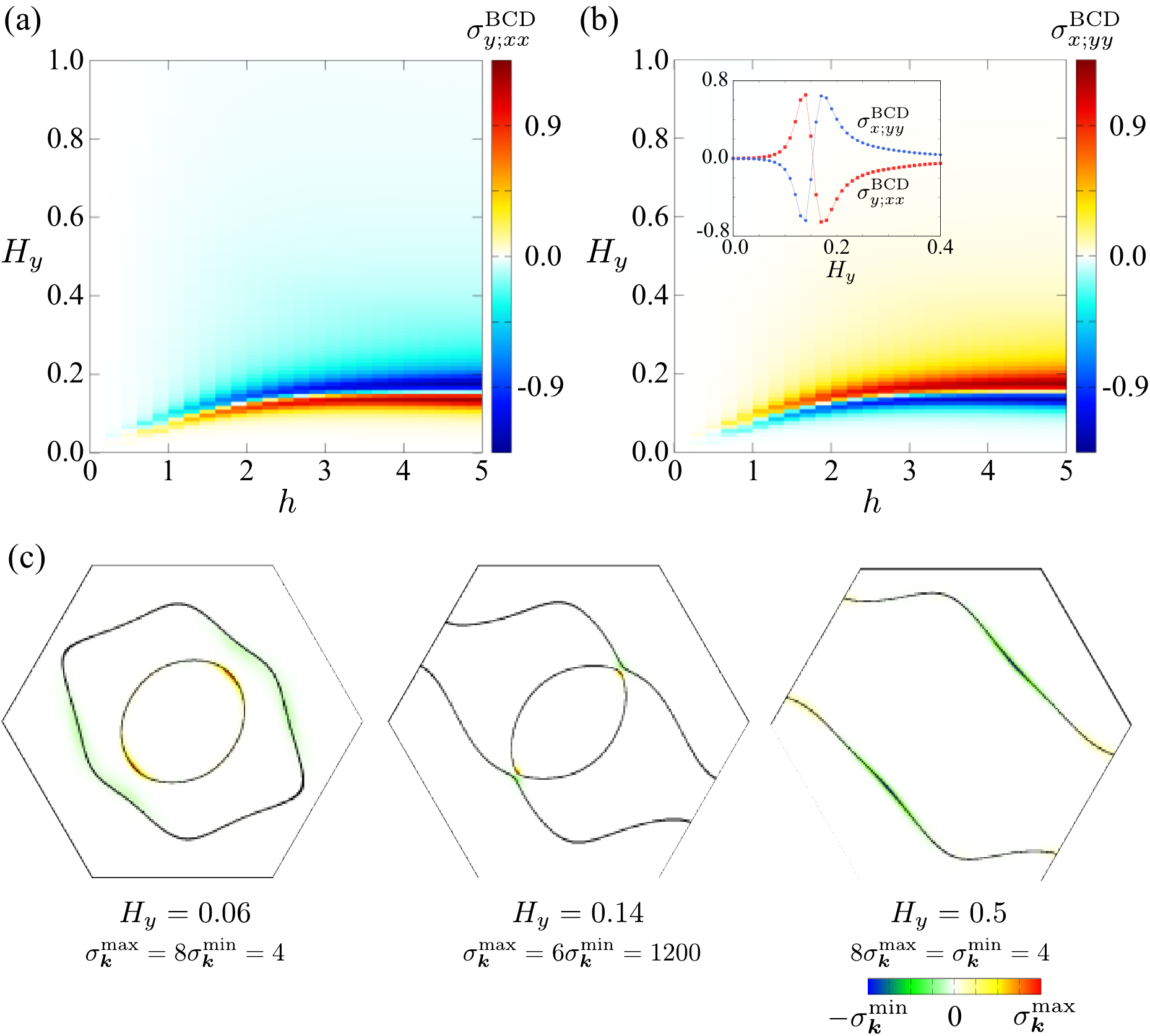} 
\caption{
\label{fig:BKL_T}
(a,b) Contour plots of (a) $\sigma^{\rm BCD}_{y;xx}$ and (b) $\sigma^{\rm BCD}_{x;yy}$ in the plane of $h$ and $H_y$ at 1/10 filling. 
The inset of (b) represents the $H_y$ dependence at $h=5$ in (a) and (b). 
(c) Contour plots of $\sigma^{\rm BCD}_{y;xx}(\bm{k})$ projected onto the Fermi surfaces for several $H_y$ at $h=5$. 
}
\end{center}
\end{figure}

On the other hand, the 120$^{\circ}$ AFM ordering in the BKL structure exhibits another nonreciprocal transport, which is referred to as the nonlinear Hall effect. 
The nonlinear Hall conductivity is calculated by 
\begin{align}
\sigma_{\eta;\mu\nu}^{\rm BCD}=\frac{e^3\tau}{2\hbar^2N_{\bm{k}}} \sum_{\bm{k},n}f_{n\bm{k}}\epsilon_{\eta\mu\lambda}D_n^{\nu\lambda}(\bm{k}) + [\mu \leftrightarrow \nu], 
\end{align}
where $D_n^{\mu\nu}(\bm{k})$ represents the BCD derived from the Berry curvature $\Omega_{n}^{\nu}(\bm{k})$; $D_n^{\mu\nu}(\bm{k})=\partial_{\mu}\Omega_{n}^{\nu}(\bm{k})$~\cite{Sodemann_PhysRevLett.115.216806}. 
In contrast to $\sigma_{\eta;\mu\nu}^{\rm D}$, $\sigma_{\eta;\mu\nu}^{\rm BCD}$ does not require the breaking of the time-reversal symmetry; the breakings of the spatial inversion symmetry and threefold rotational or mirror symmetry to activate the rank-1 electric dipole and rank-2 electric toroidal quadrupole are important~\cite{Sodemann_PhysRevLett.115.216806,kim2019prediction,Yatsushiro_PhysRevB.104.054412,Oiwa_doi:10.7566/JPSJ.91.014701}. 
Although there is no finite component of $\sigma_{\eta;\mu\nu}^{\rm BCD}$ in the TL system under any magnetic fields since such multipole degrees of freedom are not activated, it becomes finite in the BKL system owing to the active electric dipoles under the in-plane field, as discussed below. 
For example, in the case of $H_y$, nonzero components of $\sigma_{\eta;\mu\nu}^{\rm BCD}$ are given by $2\sigma^{\rm BCD}_{x;yy}=-\sigma^{\rm BCD}_{y;xy}$ and $2\sigma^{\rm BCD}_{y;xx}=-\sigma^{\rm BCD}_{x;xy}$ from the symmetry viewpoint. 

The numerical results of two components, $\sigma_{y;xx}^{\rm BCD}$ and $\sigma_{x;yy}^{\rm BCD}$, while changing $h$ and $H_y$ at $H_z=0$ are shown by the color maps in Figs.~\ref{fig:BKL_T}(a) and \ref{fig:BKL_T}(b), respectively. 
The overall feature in both cases seems to be similar; there is a sign change while changing $H_y$ for fixed $h$ and their absolute values are enhanced in the vicinity of the region where their sign change occurs, as shown in the inset of Fig.~\ref{fig:BKL_T}(b) at $h=5$. 

The enhancement of $\sigma_{y;xx}^{\rm BCD}$ is attributed to the change of the Fermi-surface topology by the band crossing. 
Figure~\ref{fig:BKL_T}(c) shows the change of the Fermi surfaces while increasing $H_y$ at $h=5$. 
We also plot the $\bm{k}$-resolved $\sigma^{\rm BCD}_{y;xx}(\bm{k})$ defined by $\sigma^{\rm BCD}_{y;xx}=\sum_{\bm{k}}\sigma^{\rm BCD}_{y;xx}(\bm{k})$ to represent the dominant contribution to $\sigma^{\rm BCD}_{y;xx}$ in momentum space. 
There are two Fermi surfaces in the low field, as shown in the case of $H_y=0.06$. 
While increasing $H_y$, the two Fermi surfaces tend to merge at two $\bm{k}$ points, where $\sigma_{y;xx}^{\rm BCD}({\bm k})$ is critically enhanced as shown at $H_y=0.14$. 
Further increase of $H_y$ leads to the decrease of $\sigma_{y;xx}^{\rm BCD}$, as the lowest band is separated from the others. 
A similar discussion holds for $\sigma_{x;yy}^{\rm BCD}$.

The origin of $\sigma_{x;yy}^{\rm BCD}$ and $\sigma_{y;xx}^{\rm BCD}$ is due to the emergence of the electric polarization induced by spin-dependent kinetic motion of electrons under the spin arrangement in the triangle unit, where the electric polarization $\bm{P}=(P_x, P_y)$ in each upward triangle is given by 
$P_x \sim  \langle\bm{s}_{\rm C}\rangle \cdot (\langle\bm{s}_{\rm A} \rangle -\langle \bm{s}_{\rm B}\rangle)$ and $P_y\sim \langle \bm{s}_{\rm C}  \rangle \cdot (\langle \bm{s}_{\rm A} \rangle+ \langle \bm{s}_{\rm B} \rangle)-2 \langle\bm{s}_{\rm A} \rangle\cdot \langle \bm{s}_{\rm B} \rangle$. 
As the effective hopping amplitudes are different depending on the bonds connected by the parallel or antiparallel spin pairs, the induced spin moments become inequivalent, which results in $\bm{P}$. 
It is noted that a similar electric polarization in the triangle unit has also been discussed in Mott insulators~\cite{Bulaevskii_PhysRevB.78.024402}.
When $H_z\neq 0$ but $H_y=0$ under the 120$^{\circ}$ spin configuration, both $P_x$ and $P_y$ become zero, which results in $\sigma_{\eta;\mu\nu}^{\rm BCD}=0$. 
Meanwhile, the in-plane magnetic field $H_y$ deforms the 120$^{\circ}$ spin configuration to have nonzero $P_x$ and $P_y$. 
As $P_x$ and $P_y$ have the same symmetry as $\sigma^{\rm BCD}_{x;yy}$ and $\sigma^{\rm BCD}_{y;xx}$, respectively, a nonzero nonlinear Hall effect appears. 
It is noted that the nonlinear Hall effect does not appear in the TL system, since both $P_x$ and $P_y$ are canceled out by considering the contributions from upward and downward triangles.

The above consideration indicates that the nonlinear Hall effect is induced even in the collinear spin configuration; it does not require the noncollinear spin configuration. 
Indeed, the $x$-spin ($y$-spin) component in $\bm{h}_i$ in addition to the uniform $y$-spin component by $H_y$ is enough to induces $P_y$ ($P_x$ and $P_y$). 
In other words, $\sigma^{\rm BCD}_{y;xx}$ still remains nonzero by dropping either of spin components, i.e., $h_{\rm A}^x=h_{\rm B}^x=h_{\rm C}^x=0$ or $h_{\rm A}^y=h_{\rm B}^y=h_{\rm C}^y=0$, 
while $\sigma^{\rm BCD}_{x;yy}$ becomes nonzero only when $h_{\rm A}^x=h_{\rm B}^x=h_{\rm C}^x=0$~\cite{comment_sigmayxx}. 
Thus, the collinear spin configuration in the triangle unit is a minimal ingredient to induce the nonlinear Hall effect in the absence of the SOC.

\begin{figure}[t!]
\begin{center}
\includegraphics[width=1.0 \hsize ]{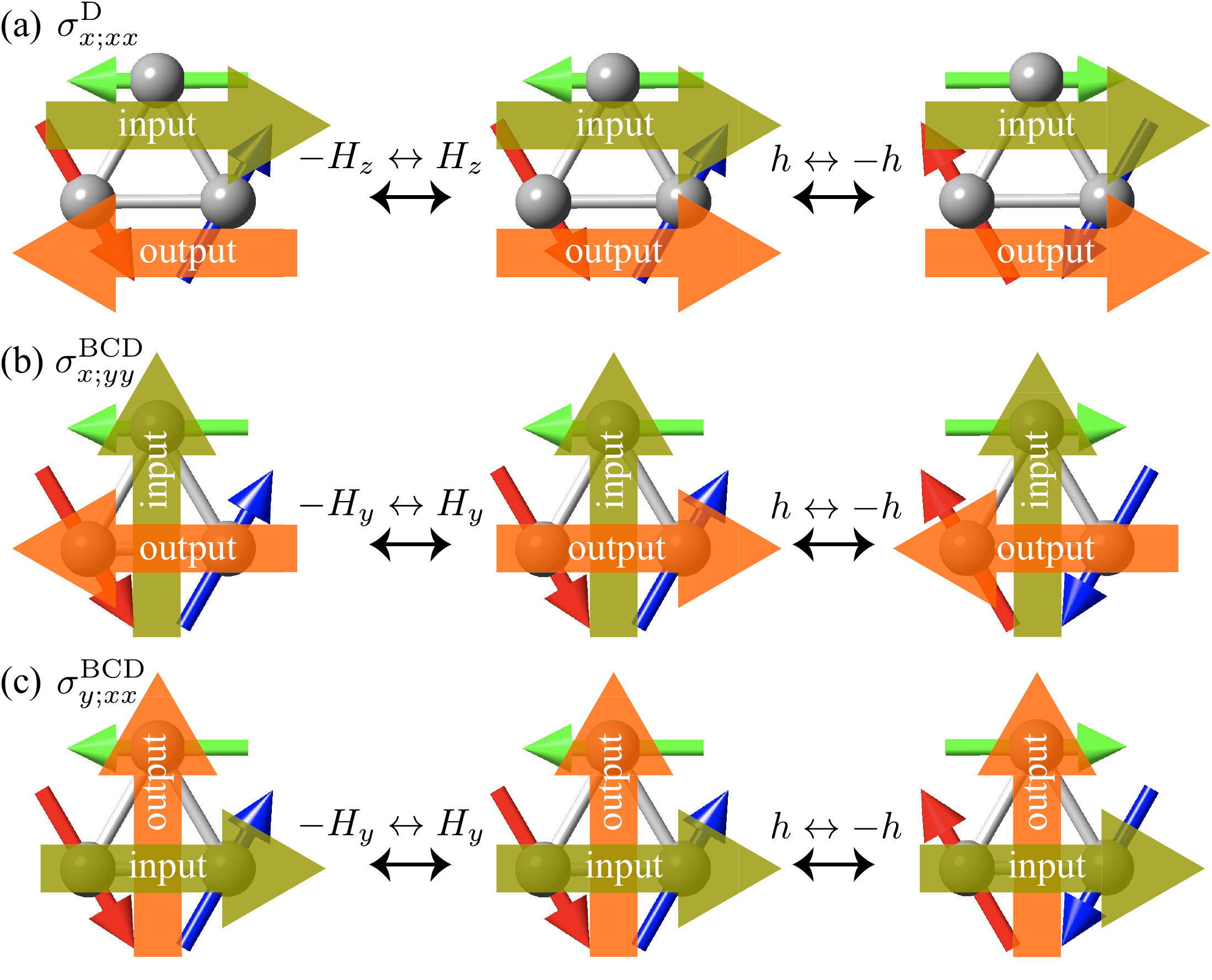} 
\caption{
\label{fig:ponti}
Schematics to represent the behaviors of (a) $\sigma^{\rm D}_{x;xx}$, (b) $\sigma^{\rm BCD}_{x;yy}$, and (c) $\sigma^{\rm BCD}_{y;xx}$ when reversing the sign of the magnetic field ($H_z$ or $H_y$) or the order parameter $h$. 
}
\end{center}
\end{figure}

\subsection{Essential model parameters}
\label{sec: Essential model parameters}

Finally, let us discuss the essential model parameters to induce $\sigma^{\rm D}_{\eta;\mu\nu}$ and $\sigma^{\rm BCD}_{\eta;\mu\nu}$ from the microscopic viewpoint. 
Although the above results indicate that both $h$ and $H_z$ (or $H_y$) are important to induce the nonlinear conductivity, their dependence is different from each other. 
To show that, we calculate the essential model parameters based on the method in Ref.~\onlinecite{Oiwa_doi:10.7566/JPSJ.91.014701}, which is given by $\sum_{ijk}C^{ijk}\sum_{\bm{k}}{\rm Tr}[v_{\eta \bm{k}}h^i(\bm{k})v_{\mu \bm{k}}h^j(\bm{k})v_{\nu \bm{k}}h^k(\bm{k})]$ for $\eta,\mu,\nu=x,y$, where $C^{ijk}$ is the model-independent coefficient, $h^i(\bm{k})$ is the $i$th power of the Hamiltonian matrix at wave vector $\bm{k}$, and $v_{\eta\bm{k}}$ is the $\eta$ component of the velocity operator; $\sigma^{\rm D}_{\eta;\mu\nu}$ ($\sigma^{\rm BCD}_{\eta;\mu\nu}$) is proportional to the real (imaginary) part of the trace.

For the Drude-type nonlinear conductivity, the essential factors of $\sigma^{\rm D}_{x;xx}$ in the TL and BKL systems are extracted as $h^2 H_z t^3$ and $h^2 H_z (t_a-t_b)$, respectively. 
This indicates that the AFM domain formation is irrelevant to $\sigma^{\rm D}_{x;xx}$, while the opposite field as $H_z \to -H_z$ reverses the sign of $\sigma^{\rm D}_{x;xx}$, as shown in Fig.~\ref{fig:ponti}(a). 
This means that the sign of $\chi_{\rm sc}$ determines the direction of the output current. 
In addition, nonzero $\sigma^{\rm D}_{x;xx}$ is induced by $t_a \neq t_b$ in the BKL system, since spatial inversion symmetry is preserved when $t_a=t_b$. 
The BCD-type nonlinear conductivities, $\sigma^{\rm BCD}_{x;yy}$ and $\sigma^{\rm BCD}_{y;xx}$ in the BKL system, show different dependences regarding $h$ and $H_y$; $\sigma^{\rm BCD}_{x;yy}$ is proportional to $h H_y (t_a-t_b)$, while $\sigma^{\rm BCD}_{y;xx}$ is $h^2 H_y^2 (t_a-t_b)$. 
In other words, $\sigma^{\rm BCD}_{x;yy}$ is reversed by reversing $H_y$ or $h$, while $\sigma^{\rm BCD}_{y;xx}$ is invariant for such a change, as shown in Figs.~\ref{fig:ponti}(b) and \ref{fig:ponti}(c), respectively. 
The difference is accounted for by the odd-parity cluster magnetic toroidal dipole in the 120$^{\circ}$-AFM structure in Fig.~\ref{fig:BKL_L}(a)~\cite{Suzuki_PhysRevB.99.174407}, which leads to a direct coupling of $\bm{P}\times \bm{H}$~\cite{Spaldin_0953-8984-20-43-434203,Hayami_PhysRevB.98.165110}. 
In the present case, $H_y$ induces $P_x$, which results in nonzero $\sigma^{\rm BCD}_{x;yy}$. 
Meanwhile, $\sigma^{\rm BCD}_{y;xx}$ is induced by a secondary effect according to the symmetry lowering under $H_y$. 
Nevertheless, the magnitude of $\sigma^{\rm BCD}_{y;xx}$ is comparable to $\sigma^{\rm BCD}_{x;yy}$ owing to the large enhancement by the band-crossing effect, as discussed in Fig.~\ref{fig:BKL_T}.

\section{Summary}
\label{sec: Summary}

To summarize, we have investigated nonlinear nonreciprocal transport in AFMs without the SOC based on the microscopic model calculations. 
We clarified that the noncoplanar spin configuration (local spin scalar chirality) causing the asymmetric band modulation induces the Drude-type nonlinear transport, while the collinear spin configuration in the triangle unit induces the BCD-type nonlinear transport under an in-plane magnetic field. 
We showed that the 120$^{\circ}$-AFM spin configurations in the TL and BKL systems are promising candidates to exhibit the nonlinear nonreciprocal transport in the external magnetic field. 
We also presented the essential model parameters in inducing each nonlinear conductivity. 

The candidate materials in the present mechanism are $X$MnO$_3$ ($X =$ Y, Sc, and Ho)~\cite{Munoz_PhysRevB.62.9498,munoz2001evolution,brown2006neutron}, BaCoSiO$_4$~\cite{ding2021field}, and Ba$_3$Mn$X'_2$O$_9$ ($X'=$ Sb and Nb)~\cite{doi2004structural,Lee_PhysRevB.90.224402}, where the noncollinear/noncoplanar AFM structures without the spatial inversion symmetry were observed.
In addition to the conventional 120$^{\circ}$ AFM structures, it is expected that the Drude-type nonlinear transport appears in other noncoplanar magnets with the local spin scalar chirality degree of freedom, such as the multiple-$Q$ states~\cite{hayami2021phase,Hayami_PhysRevResearch.3.043158,hayami2021topological}. 
The present results open another route of nonlinear transport in magnetic materials, which stimulate further exploration of functional AFM materials irrespective of the SOC.

\begin{acknowledgments}
The authors would like to thank H. Kusunose, Y. Yanagi, and R. Oiwa for fruitful discussions. 
This research was supported by JSPS KAKENHI Grants Numbers JP19K03752, JP19H01834, JP21H01037, JP22H04468, JP22H00101, JP22H01183, and by JST PREST (JPMJPR20L8). 
Parts of the numerical calculations were performed in the supercomputing systems in ISSP, the University of Tokyo.
\end{acknowledgments}

\bibliographystyle{apsrev}
\bibliography{ref}

\end{document}